\documentclass[traditabstract]{aa} 
 
\usepackage{graphicx}
\usepackage{txfonts}
\usepackage{natbib}
\bibpunct{(}{)}{;}{a}{}{,}

\begin{document}

   \title{Massive double compact object mergers: gravitational wave sources and r-process-element production sites
         }

   \author{N. Mennekens
           \and D. Vanbeveren
          }

   \institute{Astrophysical Institute, Vrije Universiteit Brussel, Pleinlaan 2, 1050 Brussels, Belgium\\
              \email{nmenneke@vub.ac.be, dvbevere@vub.ac.be}
             }

   \date{Received 3 July 2013 / Accepted March 2014}

  \abstract
   {With our galactic evolutionary code that contains a detailed intermediate mass and massive binary population model, we study the temporal evolution of the galactic population of double neutron star binaries, mixed systems with a neutron star and black hole component and double black hole binaries. We compute the merger rates of these relativistic binaries and we translate them into LIGO II detection rates. We demonstrate that accounting for the uncertainties in the relation `initial mass-final mass' predicted by massive close binary evolution and due to the possible effect of large stellar wind mass loss during the luminous blue variable phase of a star with initial mass larger than 30-40 M$_{\odot}$ and during the red supergiant phase of a star with initial mass smaller than 30-40 M$_{\odot}$ when such a star is a binary component, the double black hole merger rate may be very small, contrary to predictions made by other groups.

Hydrodynamic computations of r-process chemical yields ejected during the relativistic binary merger process have recently become available. With our galactic code that includes binaries it is then straightforward to calculate the temporal galactic evolution of the r-process elements ejected by these mergers. We conclude that except for the earliest evolutionary phase of the Galaxy ($\sim$ the first 100 Myr) double compact star mergers may be the major production sites of r-process elements and it is probable that the mixed systems dominate this production over double neutron star binary mergers.
}

            
   \titlerunning{Massive double compact object mergers}
   
   \authorrunning{Mennekens \& Vanbeveren}

   \maketitle

\section{Introduction}

The evolution of massive close binaries leading to the formation of massive double compact objects (double neutron star binaries = NS+NS, black hole-neutron star binaries = BH+NS or NS+BH\footnote{We make a distinction between BH+NS and NS+BH binaries because the progenitor evolutionary scenario is quite different.}, and double black hole binaries = BH+BH), has been studied intensively over the last four-five decades. The relatively recent and extended review papers (in review journals) of Vanbeveren et al. (1998a) and Langer (2012) may give a general state of affairs (see also the monograph `The Brightest Binaries' of Vanbeveren et al., 1998b). Theoretical population studies of massive binaries in general, massive double compact objects (we will use the abbreviation MDCO) in particular obviously rely on the evolution of massive close binaries, and therefore uncertainties in the physics that govern the evolution may imply important uncertainties in the MDCO population. To illustrate, the population studies performed with our code (= the Brussels code, De Donder and Vanbeveren, 1998, 2004a, b) predict almost no BH+BH or NS+BH mergers whereas in the simulations of Voss and Tauris (2003) the double BH mergers are about two orders of magnitude larger than the BH+NS and NS+NS mergers. Dominik et al. (2012) used the StarTrack\footnote{The StarTrack population code has been described by Belczynski et al. (2008) and references therein.} population code to estimate the MDCO population and they confirm the Voss and Tauris findings. Therefore, one of the purposes of the present paper is to investigate where the differences with the Brussels predictions may come from. Moreover, since the year 2004 (the year where we published our last MDCO population studies), new insights in the physics of massive close binary evolution have emerged. The present paper highlights how this new physics is implemented in our population code and how this affects the MDCO-population prediction. To make a comparison with other simulations (e.g., Voss and Tauris, 2003, Dominik et al., 2012, Belczynski and Dominik, 2012\footnote{This paper is a preliminary version of Belczynski et al. (2014), the latter which appeared during the reviewing process of the present work.}), we will also present our results in terms of detection rates of the advanced laser interferometer gravitational wave observatory LIGO II that will probably be operational in 2015.

Stellar evolution in general, the evolution of massive stars and binaries in particular depends on the chemical composition. Progenitors of NS+NS, NS+BH, BH+NS or BH+BH binaries observed at present may have been formed when the galactic metallicity was lower. In order to estimate the present population of double compact star binaries, it may therefore be indispensable to link a binary population code and a galactic chemical evolutionary code where the evolution of the chemical yields is consistently computed by using realistic single star and binary yields as function of metallicity.

Most of the galactic chemical evolutionary codes that exist today ignore the effects of binaries although there is increasing evidence that at least 50\% of the intermediate mass and massive stars have a companion that is close enough to affect significantly its evolution (Kouwenhoven et al., 2007; Sana et al., 2013). Between 1998 and 2004 the Brussels group has investigated systematically the effects of binaries on galactic chemical evolution (for a review see De Donder and Vanbeveren, 2004, further noted as DDV04). In particular, a full intermediate mass close binary population code was linked to a galactic model in order to compute in detail the SNIa rate and the effect on the age-metallicity relation (the model to calculate the SNIa rate has been updated recently by Mennekens et al., 2010, 2013). Moreover, DDV04 also implemented a detailed massive close binary population code in their galactic model. In this way it was possible to follow the MDCO-merger rate as function of galactic age (i.e. as function of metallicity by using the age-metallicity relation that was consistently calculated accounting for the presence of binaries). It was suspected that MDCO-mergers could be important r-process formation sites but detailed hydrodynamic ejecta computations were still very rare (Oechlin et al., 2002). In 2004 we could only compare the predicted temporal evolution of the MDCO-merger rate with the observed galactic temporal evolution of r-process elements (europium in particular). Although a quantitative comparison could not be made at that time, we concluded that within the uncertainties of binary population synthesis the predicted shape of the temporal evolution of the merger rate is comparable to the observed temporal pattern of galactic europium for [Fe/H] $> -2$, but the early galactic evolution is not well covered (see also Argast et al., 2004). Since then, hydrodynamic ejecta have been computed for a set of MDCO-mergers (Korobkin et al., 2012) and it becomes feasible to calculate more realistically the temporal evolution of r-process elements produced by MDCO-mergers. This is a second major purpose of the present study.

The paper is organized as follows. The single star/binary population and galactic code that we use here relies on the original code as it has been described in DDV04. Section 2 summarizes the basic ingredients of this 2004-version. Section 3 deals with all the updates since 2004. The MDCO-merger-population and the r-process enrichment due to MDCO-mergers are discussed in Sect. 4.

\section{The Brussels massive single star/binary population and galactic codes: the 2004-versions}

As stated in the introduction a detailed description of the Brussels population/galactic code (which includes binaries) is published in DDV04. Here we give a short summary. 

\subsection{Massive single star evolution}

The Brussels population/galactic code uses the results computed with the Brussels stellar evolutionary code that has been described in Vanbeveren et al. (1998a, b). This code adopts the opacities from Iglesias et al. (1992) and the nuclear reaction rates from Fowler et al. (1975). Semi-convection is treated according to the criterion of Schwarzschild and Harm (1958) and convective core overshooting is included as in Schaller et al. (1992).

The formalisms used to calculate the effect on evolution of stellar wind mass loss rates deserve some special attention. We can distinguish between the rates of core-hydrogen-burning (OB-type) stars prior to the Luminous Blue Variable (LBV) phase, the LBV-mass loss rates, the red supergiant (RSG) rates and the rates of hydrogen deficient core-helium-burning (Wolf-Rayet = WR type) stars.

For OB-type stars we used the formalism of De Jager et al. (1988) but we remark that as far as population and/or galactic studies is concerned, more recent formalisms like the one proposed by Vink et al. (2000) or Pauldrach et al. (2012) give very similar results. Even more, if the OB-type star rates that are presently observed will be confirmed, it can be concluded that for overall (integrated) massive star population/galactic evolution, the importance of OB-type star stellar wind mass loss is marginal.

We will demonstrate that the LBV-phenomenon in general, LBV-wind mass loss in particular is very important for the results and the conclusions of the present paper. For a throughout review on LBVs we like to refer to the monograph `Eta Car and the Supernova Impostors' (Davidson and Humphreys, 2012), here we list some facts which are important for the scope of the present paper (see also the review paper of Vink, 2009). LBVs have a luminosity Log L/L$_{\odot} >$ 5.4-5.5 and they occupy a region in the HR-diagram that corresponds to the hot part (= the beginning) of the hydrogen shell burning phase of stars with an initial mass larger than 25 M$_{\odot}$. The observed LBV-mass loss rates are very large and range between $10^{-5}$ and $10^{-3}$ M$_{\odot}$/yr. On top of that LBVs are known to experience major mass eruption phases like the famous Eta Car eruption in the 19th century where almost instantaneously about 10 M$_{\odot}$ were expelled by a process that is as yet unclear. The observations reveal that the region in the HR-diagram which is located to the right of the region occupied by the LBVs with a luminosity Log L/L$_{\odot} >$ 5.5-5.6 (corresponding to stars with an initial mass larger than 30-40 M$_{\odot}$) is almost void. There are a few yellow hypergiants but no red supergiants. This is also observed in the Magellanic Clouds. Based on this fact and on the observed mass loss rates it was suggested by Humphreys and Davidson (1979, 1984) that the LBV-mass loss has to be so large that it prohibits a hydrogen shell burning star with initial mass $>$ 30-40 M$_{\odot}$ from evolving into a red supergiant, and that this scenario is metallicity independent. From evolutionary point of view this is a very robust criterion to calculate the LBV-mass loss rate that is needed. This was investigated by the Geneva stellar evolution team (Schaller et al., 1992) who indeed confirmed the suggestion of Humphreys and Davidson. The evolutionary calculations then reveal that stars with an initial mass $>$ 30-40 M$_{\odot}$ lose their hydrogen rich layers by LBV mass loss and they become Wolf-Rayet stars without becoming RSGs first. In Sect. 2.2 we explain how this affects the evolution of binaries with component masses $>$ 30-40 M$_{\odot}$ (the LBV scenario of massive binaries as it was introduced by Vanbeveren, 1991).

Massive stars with an initial mass $\leq$ 30-40 M$_{\odot}$ become RSGs and their further evolution is therefore governed by RSG stellar wind mass loss. Until very recently, most of the research groups that studied massive single star evolution used the RSG mass loss rate formalism of De Jager et al. (1988). However, Vanbeveren et al. (1998a, b) argued that the RSG rates may be significantly larger than those predicted by the De Jager et al. formalism and this has been confirmed recently by Van Loon et al. (2005). Since 1998, all our massive star evolutionary tracks have been computed using these larger RSG rates. Therefore, since all our population/galactic studies published since 1998 rely on these tracks, they correspond to massive star evolution that accounts for larger RSG mass loss rates. Note that other groups that study massive single star evolution who used De Jager et al. rates for RSGs have recently adapted their evolutionary code and also use larger RSG rates (Georgy, 2012; Georgy et al., 2012; Ekstrom et al., 2012). One of the main consequences of larger RSG mass loss rates is that stars with an initial mass between 15-20 M$_{\odot}$ and 40 M$_{\odot}$ lose their hydrogen rich layers during the RSG phase. This obviously affects the population computations that predict the number of WR single stars (Vanbeveren et al., 1998c, 2007; Sander et al., 2012). In Sect. 2.2 we will discuss how this RSG stellar wind mass loss affects the evolution of massive binaries with component masses $\leq$ 30-40 M$_{\odot}$.

The mass loss rate during the WR phase is of prime importance. In our evolutionary code we use the formalism that was discussed in Vanbeveren et al. (1998a, b, c) but note that as far as the effect on stellar evolution is concerned our formalism predicts results that are very similar to the results when the formalism of Nugis and Lamers (2000) is used.

Finally, all the mass loss rates (except the LBV rates) are assumed to be metallicity (Z) dependent (we used a $\sqrt{Z}$ dependency) and this obviously affects the final mass of the star and/or the mass of the compact star remnant (neutron star or black hole).

\subsection{Massive close binary evolution}

Prior to the Roche lobe overflow (RLOF) it is assumed that the components of massive binaries evolve as massive single stars and therefore the discussion of subsection 2.1 also applies here. Following the classification of Kippenhahn \& Weigert (1967) and Lauterborn (1970), we distinguish three main phases of RLOF, which correspond to the three major expansion phases during stellar evolution: case A where the RLOF takes place during the core hydrogen burning phase of the mass loser, case B where the RLOF occurs during the hydrogen shell burning phase prior to the blue loop phase during central helium burning (see also Schaller et al., 1992), and case C where the RLOF begins after helium has been depleted in the core. Case B RLOF is additionally divided into early case B or case Br, where at the onset of RLOF the envelope of the mass loser is mostly radiative, and late case B or case Bc, where the primary has a deep convective envelope at the beginning of the RLOF phase. In case A and case Br, mass loss will cause the donor to shrink within its Roche lobe again. Mass transfer will thus be dynamically stable, and result in a RLOF phase which may or may not be conservative. The only exception is when the mass ratio of the system is lower than 0.2, in which case the Darwin instabililty will eventually cause the mass transfer to become unstable and result in a common envelope phase. In case Bc and case C, however, mass loss will lead to a radius increase, and mass transfer will quickly become unstable, also resulting in a common envelope phase. The exact boundary between stable case Br and unstable case Bc mass transfer is determined as follows: from the donor ZAMS mass, the theoretical radius is computed at the time these star's outer layers become deeply convective. If the orbital period of the system is sufficiently small that mass transfer has started before the donor reaches this radius, mass transfer will be stable. A star that already went through a first phase of RLOF during hydrogen shell burning may fill its Roche lobe for a second time during helium shell burning and undergo case BB RLOF (Delgado \& Thomas, 1981).

DDV04 describe the physics of the different cases and how this is implemented in our population code. A recent summary was given in Vanbeveren et al. (2012). Although the latter paper deals with intermediate mass binaries, as far as the RLOF is concerned the majority of these basic ingredients remain valid also for massive stars and will not be repeated here. We like to remind the reader that during the evolution of a binary frequently a situation is encountered where both components merge as a consequence of RLOF/CE/spiral-in. Our merger criterion has been discussed in the two papers cited above. Below we summarize some ingredients of our code that are important in order to better understand the results of the present paper.

1.	The stellar wind mass loss (during the OB-phase, the LBV, RSG and WR phases) is assumed to be (spherically) symmetric and consequently it results in an orbital period increase that is described by the formula
			
\begin{equation}
\frac{P_f}{P_i} = \left( \frac{M_{1f}+M_{2f}}{M_{1i}+M_{2i}} \right)^{-2}
\end{equation}

2.	In a case C massive binary the components with an initial mass $\leq$ 30-40 M$_{\odot}$ go through a RSG phase prior to the onset of RLOF and therefore one has to account for the effect of the RSG stellar wind mass loss rates on the RLOF. As discussed in subsection 2.1 the RSG wind removes most of the hydrogen rich layers of stars with Solar type metallicity and with an initial mass between 15-20 M$_{\odot}$ and 30-40 M$_{\odot}$ and therefore in binaries with a component with a mass in this range case C RLOF is suppressed. This binary scenario has been originally discussed by Vanbeveren (1996) and Vanbeveren et al. (1998a) and has been further referred as the `RSG scenario of massive close binaries'. Accounting for the recent redetermination of the mass loss rates of RSGs of Van Loon et al. (2005), the latter RSG binary scenario becomes very plausible.

3.	As discussed in subsection 2.1 it is conceivable that LBV stellar wind mass loss suppresses the redward evolution during hydrogen shell burning of stars with an initial mass larger than 30-40 M$_{\odot}$ and that stars in this mass range lose most of their hydrogen rich layers by this LBV mass loss process. As a consequence it cannot be excluded that the LBV mass loss rate suppresses the RLOF/common envelope phase in case Br/case Bc/case C binaries when the mass loser has a mass larger than 30-40 M$_{\odot}$ (the LBV scenario of massive binaries as it was introduced by Vanbeveren, 1991). In these binaries the orbital period variation satisfies equation (1). We will demonstrate that this assumption affects critically the predicted population of double-BH-binary mergers.\footnote{At first glance, the period increase due to LBV winds may seem to preclude the formation of (observationally found) massive close binaries with black hole components, such as Cyg X-1 (5.6 d) and IC10 X-1 (1 d). However, (a combination of) different effects can still result in such systems being predicted in our standard model. First remind that the LBV scenario does not apply to case A binaries and such progenitor scenario cannot be excluded for these two X-ray systems. Moreover, when the system undergoes a supernova explosion the resulting kick velocity (see point 6) can also decrease the orbital separation. Furthermore, the latter system contains a WR star with a minimum mass of 17 M$_{\odot}$ (Prestwich et al. 2007), and may hence have gone through a common envelope phase. As an example, a system initially consisting of 50 and 38 M$_{\odot}$ stars with an initial orbital period of 20-200 days, will lead to an LBV phase. This and the ensuing supernova explosion of the primary will create a range of possible orbital periods. A subset of these will allow for the initiation of a CE phase by the secondary (which is less massive than 40 M$_{\odot}$), decreasing the orbital period and leading to a system like IC10 X-1.} Studies such as Vink \& de Koter (2002) arrive at LBV mass loss rates of 10$^{-5}$-10$^{-4}$ M$_{\odot}$/yr, but these are valid for the steady mass loss in dormant phases. For our purposes, one should use the average of these and the much more significant mass loss rates during eruptions (the most famous example of course being $\eta$ Car, which lost some 10 M$_{\odot}$ in one episode in the 19th century, giving it an average mass loss rate of 10$^{-1}$ M$_{\odot}$/yr during that century). The results of the present paper with the label `LBV on' are calculated assuming an average LBV wind (the average of the eruption + inter-eruption mass loss) of 10$^{-3}$ M$_{\odot}$/yr. Notice that the results presented here would be entirely similar when other average values would be adopted provided that these values are large enough so that they prevent the occurrence of the RLOF.

4.	Despite 4 decades of binary research it remains unclear whether or not the stable Roche lobe overflow in massive case Br binaries is conservative. Therefore, it is necessary to investigate the effect on population synthesis results of different accretion efficiency characterized by $\beta$, which is the fraction of matter lost by the donor that is accreted by the gainer ($\beta$= 1 then corresponds to the conservative case).  When mass is lost from the system, it is necessary to make an assumption about how much angular momentum this lost mass takes with. This quantity is obviously dependent on the physical model of how this matter escapes from the system, and is critical for the orbital period evolution during the mass transfer phase. Our standard model assumes that matter will escape via the second Lagrangian point L$_2$, from where it forms a circumbinary disk (van den Heuvel, 1993). A ``bare-minimum'' for the radius of this disk is obviously equal to the distance from L$_2$ to the center of mass. However, Soberman et al. (1997) concluded that circumbinary disks are stable (e.g., the matter in the disk will not have the tendency to fall back towards the binary) only when their radii are at least 2.3 times the binary separation. This disk model with disk radius equal to 2.3 times the binary separation is standard in the Brussels binary population code. How the binary period varies then has been described in DDV04. Notice that with this formalism mass lost from the binary implies significant orbital angular momentum loss resulting in a binary period decrease increasing the probability that the two binary components merge. As will be discussed in Sect. 4 the adopted angular momentum loss model significantly affects the predicted number of double NS binaries and mergers.

5.	After case B mass transfer, the remnant helium star may grow to giant dimensions during helium shell burning and fill its Roche lobe again, initiating a phase of case BB mass transfer (Habets, 1986a, b; Avila Reese, 1993). In DDV04 we discuss in more detail this mass transfer process. For the scope of the present paper it may be important to know that when the mass gainer is a normal star then case BB mass transfer is assumed to proceed in a conservative way, whereas if the mass gainer is a NS or a BH, case BB mass transfer is assumed always to result into a common envelope evolution where we use the values for the parameters $\alpha$ and $\lambda$ in the common envelope formalism (we use the formalism of Webbink, 1984) which are similar as those for a case Bc/C common envelope.

6.	The evolution of a massive binary up to the formation of a double compact star binary obviously depends critically on how an asymmetric supernova explosion of one component affects the orbital parameters of the binary. As a consequence of an asymmetric supernova explosion the remnant NS or BH receives a kick velocity and this allows us to compute the effect on binary parameters in a straightforward way. The kick velocity distribution is linked to the observed pulsar velocity distribution. We described the latter with a $\chi^2$-distribution with average velocity 450 km/s. Other authors sometimes use a Maxwellian distribution with a similar average (e.g. following Hobbs et al. 2005) but it can readily be understood that both treatments will yield very similar results. Some recent results however point to lower velocities (e.g. Belczynski et al. 2010b). To test the influence on our results, a calculation will be presented with an average kick velocity of 265 km/s. When the binary survives the SN explosion the post-SN system is in most of the cases eccentric. Our population code does not account for tidal effects in binaries but we assume that eccentric binaries after the first SN explosion become circularized before the collapse of the other star. Of course, when the binary survives the second SN explosion the resulting eccentricity is essential in order to calculate the merging timescale of the two relativistic stars.

\begin{figure}[t]
\centering
   \includegraphics[width=8.4cm]{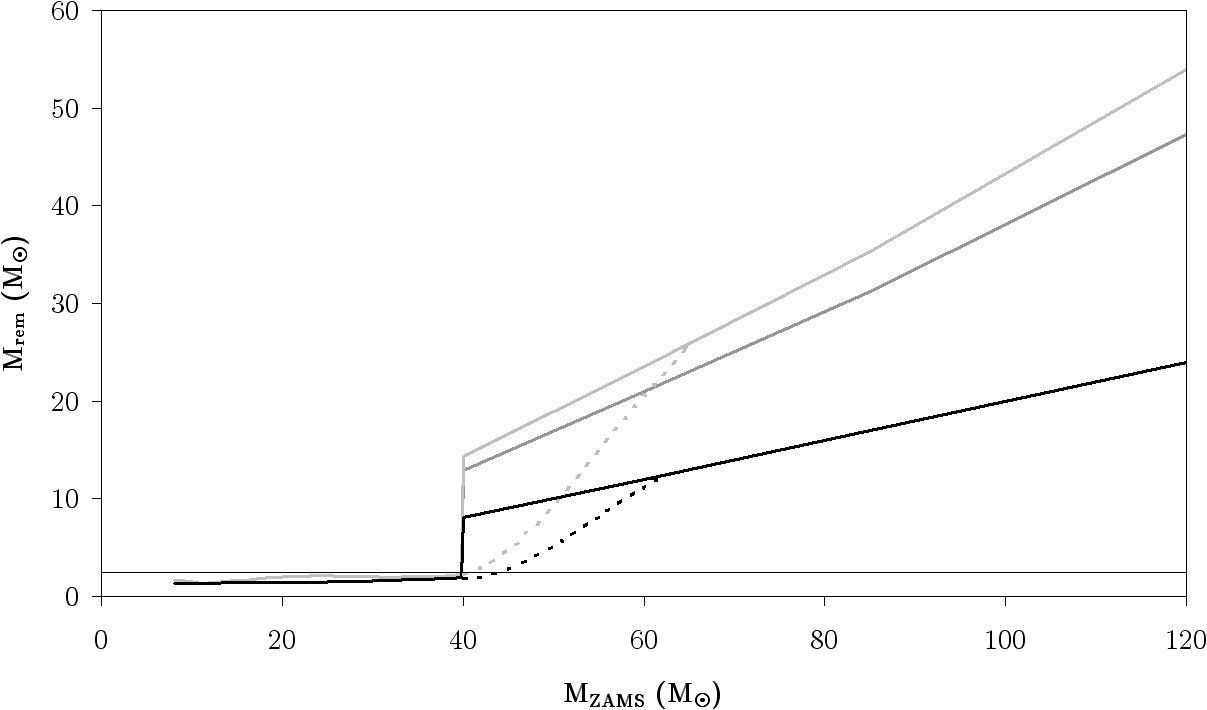}
     \caption{Brussels initial mass - remant mass relation for a donor in a binary star, for Z=0.02 (black), Z=0.006 (dark gray) and Z=0.002 (light gray). Dashed sections are the relations used to avoid the discontinuity at 40 M$_{\odot}$ (see text).}
     \label{fig:MiMf}
\end{figure}

7.	Particularly important for the results of the present paper is the initial-final mass relation of massive binary components. The relation for mass losers of massive binaries that is implemented in our population code is shown in Fig. \ref{fig:MiMf} and has been discussed in detail in DDV04. To summarize this relation relies on detailed massive binary evolutionary computations that account for the stellar wind mass loss phases as discussed above. To link the final CO-core at the end of core helium burning and the mass of the final Fe-core we use the model B explosion model of Woosley and Weaver (1995). The apparent discontinuity at 40 M$_{\odot}$ is due to the fact that it is assumed that binary components with an initial mass $\geq$ 40 M$_{\odot}$ at the end of their life collapse directly into a black hole without a SN explosion. To explore the consequence of this assumption we also calculated models where the final black hole mass for stars with an initial mass $\geq$ 40 M$_{\odot}$ follows the continuous relation. The difference in mass is then expelled during a SN explosion where we use the model description of Fryer et al. (2012). The authors discuss two models: the `Rapid' and the `Delayed' supernova mechanism. Both mechanisms are implemented in our code and we will discuss the population differences.

8.	In order to check the validity of our assumptions, we can make a comparison with another code. It is however not always clear what is the initial-final mass relation used by other groups. Let us focus on the population synthesis code StarTrack used by Dominik et al. (2012) and Belczynski \& Dominik (2012). The StarTrack code is based on the single star relations shown in Fig. 10 of Fryer et al. (2012). In codes such as this, making use of the Hurley et al. (2000) formalism, the detailed single star calculations are then adapted to include binary effects through a wind-like process, outlined by e.g. (Sect. 5.6 of) Belczynski et al. (2008). As the detailed results of these have however not been published in e.g. graphical evolution tracks, it is very difficult to judge how strongly these adapted tracks deviate from the single star tracks. Another difference is in the stable mass transfer and merger criterions. Our assumptions are described in DDV04. They basically require the initial mass ratio of case A/Br systems to be above 0.2 for stable mass transfer, and both post-RLOF components to be confined by their Roche lobes in order for the system not to merge. StarTrack uses different assumptions, which require a much higher initial mass ratio ($\sim$ 0.6-0.65) for mass transfer to be stable and a likely merger to be avoided (see Toonen et al. 2014 for a comparison between four different intermediate mass binary population synthesis codes). Furthermore, the angular momentum loss assumptions in StarTrack are different. Whereas we assume that mass is lost with the specific orbital angular momentum of L$_2$, StarTrack makes an assumption that is very similar to gainer orbital angular momentum loss. Anticipating, in Sect. 4 we will show that when our assumptions are changed to correspond as much as possible to those of StarTrack (but with the Hurley et al. 2000 initial-final mass relation for single stars), the results correspond very well. This is an encouraging indication (as is Toonen et al. 2014 for the intermediate mass range) for the usefulness of binary population synthesis approaches. However, it also illustrates that the procedure used by Fryer et al. (and other users of the Hurley et al. 2000 analytical formalisms) to adapt these results to binaries might not change the evolution tracks sufficiently for them to correspond well to our detailed binary evolutionary tracks.

\subsection{The galactic code}

In the 2004 version of the galactic code we used the two-infall-galaxy-formation model of Chiappini et al. (1997) and the star formation prescription of Talbot \& Arnett (1975) that was adapted to the infall model by Chiosi (1980). With the Brussels binary evolutionary code we first calculated single star yields accounting for the OB, LBV/RSG and WR stellar wind mass loss rates discussed in subsection 2.1. We then computed in detail the binary chemical yields\footnote{At this stage we restricted ourselves to the $\alpha$-elements.} as function of primary mass, binary mass ratio and binary period. All yields were computed as function of metallicity. These yields together with a full binary population code were linked to the galactic model. In particular, the binary population code also computes the intermediate mass binary population and it was therefore straightforward to calculate the temporal evolution of the SN Ia rate (and the resulting Fe-enrichment) in detail, either adopting the single degenerate SNIa scenario (Hachisu et al., 1999), the double degenerate scenario (Webbink, 1984; Iben \& Tutukov, 1984) or both scenarios together.

\section{The Brussels massive single star/binary population and galactic codes: the present-versions}

\begin{figure}[t]
\centering
   \includegraphics[width=8.4cm]{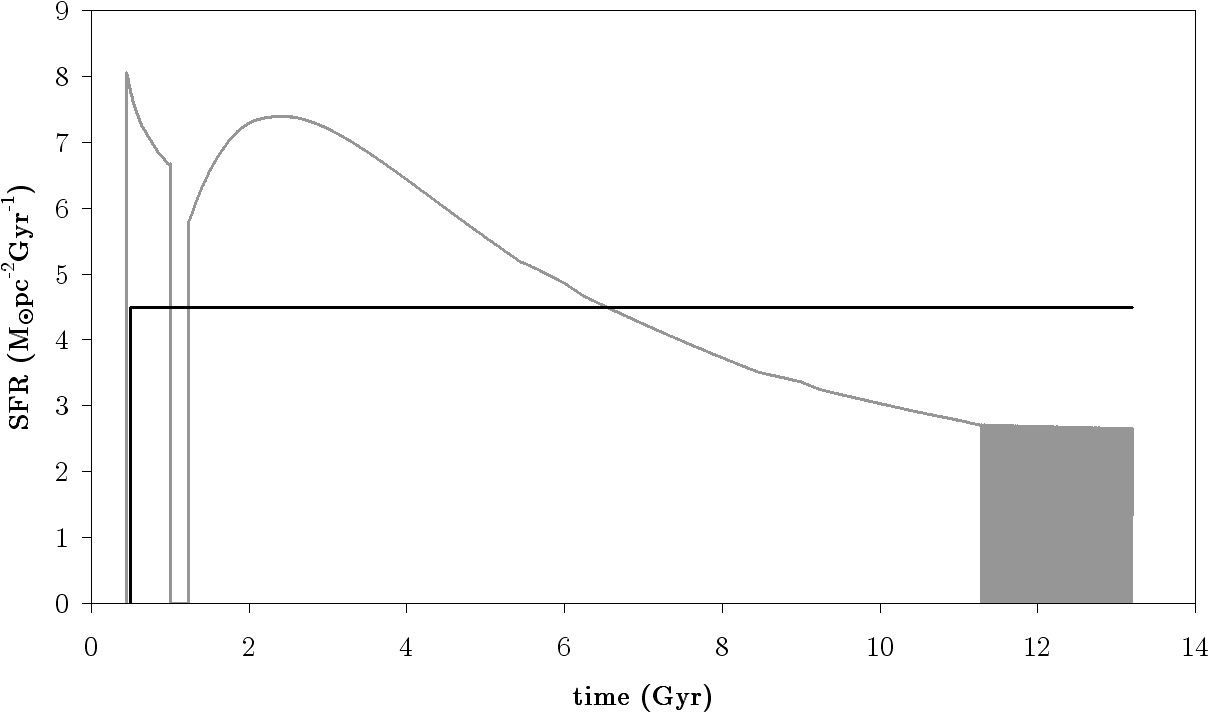}
     \caption{Star formation rate as a function of time obtained with the current version of the two-infall model (denoted SFR$_1$ in the text and tables; gray) vs. constant SFR (SFR$_2$; black).}
     \label{fig:SFR}
\end{figure}

Since 2004, a number of major updates have been implemented in our code(s).

a.	The common envelope (CE) phase: to estimate the response of a binary to the common envelope phase in case Bc/C binaries, we use the expression describing the energy balance of the CE described by Webbink (1984). The expression contains two parameters: $\alpha_{CE}$ is defined as the efficiency of the transfer of orbital energy into escape energy of the common envelope ($0 < \alpha_{CE} < 1$) and $\lambda$ which is a measure of the binding energy of the envelope of the mass loser. In the binary population simulations performed by some groups both parameters are assumed constant and independent from the masses and period (which determines the radius of the mass loser at the onset of the CE) of the binary. Studies such as Dewi \& Tauris (2000), Xu \& Li (2010) and Loveridge et al. (2011) give self-consistently calculated values for $\lambda$ as a function of stellar mass and radius. Given the significant remaining uncertainties on them we implement these results in a straightforward way. One of these uncertainties is the difference between the calculations under the assumption that the binding energy of the donor consists only of its gravitational energy, or of its entire internal energy (termed $\lambda_g$ and $\lambda_b$ respectively). For stars below 10 M$_{\odot}$, we implement the results of Dewi \& Tauris (2000), and for more massive stars those for Z=0.02 of Xu \& Li (2010). In both cases, an average of the author's $\lambda_g$ and $\lambda_b$ is used. To obtain interpolation points for stars more massive than 20 M$_{\odot}$, we make use of the Xu \& Li results communicated by Dominik et al. (2012) in their Fig. 3 and 4. The slow variation of $\lambda$ with masses in this range makes extrapolation for M$>$60 M$_{\odot}$ quite reliable. No metallicity effect is taken into account, as the difference between values for Z=0.02 and Z=0.001 is much smaller than the difference between $\lambda_g$ and $\lambda_b$ for a single value of Z. It should of course also be kept in mind that this determination for $\lambda$ does not solve the $\alpha$-uncertainty. Therefore, all calculations are presented for values of $\alpha$ ranging from 1.0 down to 0.1. Ivanova \& Chaichenets (2011) use enthalpy considerations to find that in reality $\lambda$ may be 2-5 times higher than the values discussed above. To reflect these results, we also do a calculation where $\lambda$ is taken five times higher than its calculated value. Combined with $\alpha=1$, this provides an absolute upper limit on the CE efficiency.

b.	In the 2004-version of the binary population code the RLOF in case A binaries was treated as in case Br. However, by definition, a case A binary has a smaller period than a case Br. This means that during the RLOF in case A binaries the gas-stream from the mass loser will hit directly the mass gainer (instead of first forming a Keplerian disk which is the case in case Br binaries) and this favors conservative mass transfer (for a review see Langer, 2012). In all our simulations case A is treated conservatively ($\beta$ = 1)\footnote{$\beta$ is defined as the amount of mass lost by the mass loser during its RLOF that is accreted by the mass gainer.}. When we present simulations with $\beta < 1$ the latter value refers to the amount of mass lost by the loser and accreted by the gainer in case Br binaries. Note that with our merger criterion (DDV04, Vanbeveren, 2012) $\sim$20-30\% of all case A binaries merge. The percentage of case Br mergers depends on the assumed value of $\beta$ (see also Vanbeveren et al., 2013) and this will obviously affect the predicted population of double compact star binaries.

c.	Population studies of binaries rely on the adopted binary period distribution. Most research groups use a distribution that is flat in the Log (also we did that in 2004). However, Sana et al. (2013) investigated binary properties of a statistically significant number of O-type stars in the Galaxy and in the Large Magellanic Cloud. They concluded that the binary frequency is very high ($\geq$ 50\%) and that the period distribution is skewed towards short period. They propose a period distribution (Log P)$^{-0.55}$. We also made simulations assuming that this period distribution applies for all massive binaries.

d.	Based on detailed evolutionary calculations of Nomoto (1984, 1987), Podsiadlowski et al. (2004) suggested that stars in binaries that develop a helium core (after case A/B Roche lobe overflow) with mass between 1.4 M$_{\odot}$ and 2.5 M$_{\odot}$ may end their life in a prompt (fast) electron-capture SN (ECSN) where the resulting NS is born with a small kick. We will present simulations with and without this ECSN channel (ECSN off or on). The case `ECSN on' means that the simulations were done assuming that the kick velocity of the NS resulting from these ECSN is zero.

e.	It is conceivable that for BH formation, kick velocities are significantly lower than for a NS (see e.g. Mirabel \& Rodrigues (2003) and the discussion by Belczynski \& Dominik 2012). This would imply that systems containing a BH (or two) are less likely to be disrupted and thus more likely to merge. To illustrate the influence thereof, we will calculate a model with kicks off for all BH formation.

f.	Fryer et al. (2012) presented a detailed study of the supernova explosion of massive stars and the formation of neutron stars or black holes. The authors discuss two models: the `Rapid' and the `Delayed' supernova mechanism. For non-ECSN, both mechanisms are implemented in our code and we will study the population differences.

g.	A simulation of the evolution of a galaxy obviously depends on the adopted model for the temporal evolution of the gas density and the star-formation rate (SFR). The method used by Dominik et al. (2012) to normalize their calculated rates to the galactic ones is very simple. They consider the Milky Way to be a galaxy with an age of 10 Gyr and a constant global star formation rate of 3.5 M$_{\odot}$/yr. Our model attempts to recreate the galactic SFR in more detail, but it also allows to be simplified to the above case. We start by recovering the SFR with a self-consistent model for the Solar neighborhood. The latter is defined as a cilindrical region of about 1 kpc centered on our Sun. As in most galactic models, we give this local SFR in units of M$_{\odot}$pc$^{-2}$Gyr$^{-1}$. This value is then converted to a global galactic SFR by taking into account the relative size of the Galaxy compared to the Solar neighborhood. That conversion factor is assumed to be 1000, motivated by the Galactic diameter of 30 kpc. This means that we implicitly assume the Solar neighborhood, halfway between the center and the edge, to be the `average' galactic environment. If the local SFR is taken to be constant and equal to the currently observed value of 4.5 M$_{\odot}$pc$^{-2}$Gyr$^{-1}$, this thus results in a global galactic SFR of 4.5 M$_{\odot}$/yr. That model can then be readily compared to the results of Dominik et al. (2012) by multiplying them with a factor of 3.5/4.5=0.78. An alternative is to use for the Solar neighborhood the SFR which follows from the two-infall model by Matteucci et al. (2009). This SFR (shown in Fig. \ref{fig:SFR} along with the constant one) shows a spike of star formation at early times and then a slow decline. In the last two Gyr, the SFR oscillates between zero and a value consistent with current (assumed to be 13.2 Gyr) observations. The total amount of star formation integrated over time is however very similar to the case of the constant SFR. This model can then also be extrapolated to the entire galaxy. Such method allows to study the influence on the merger rates of a global galactic SFR which was higher in the past than today.

\begin{figure}[t]
\centering
   \includegraphics[width=8.4cm]{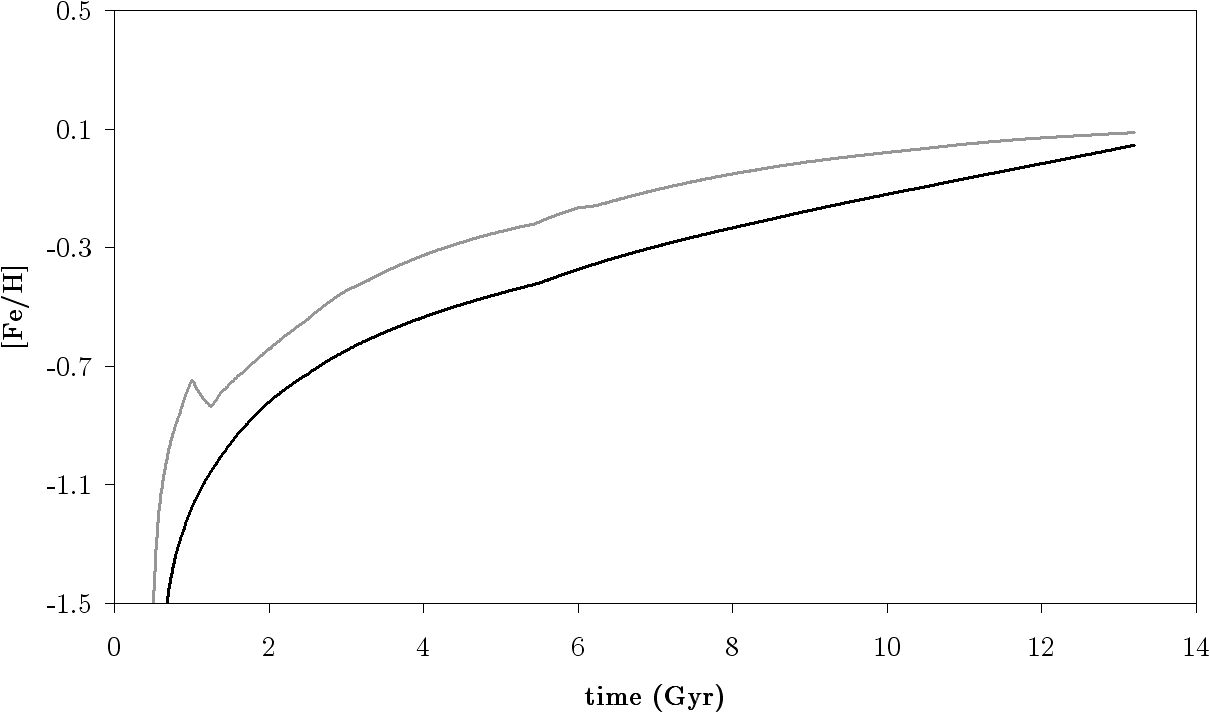}
     \caption{Age metallicity relation, i.e. [Fe/H] as a function of time, for the SFR following from the two-infall model (gray) and for a constant SFR (black).}
     \label{fig:AMR}
\end{figure}

h.	It is obvious that since MDCO merger rates depend on metallicity, one has to follow the metallicity evolution of the galaxy throughout time. The method used by Belczynski \& Dominik (2012) to account for this is very simple. They perform a calculation for Solar metallicity (Z=0.02) and one for low metallicity (Z=0.002). The final rate is then simply taken as the average of the two. We prefer to follow the evolution of Z in more detail. For a complete description of our galactic model, including binaries, we refer to DDV04. The initial Z-value of the Galaxy is assumed to be zero\footnote{The DCO formation rates, merger rates and yields for this Z are obtained by extrapolation of the results at Z=0.002. It was explained in Sect. 2.1 why we believe LBV wind mass loss rates to be Z-independent, and thus also large at low Z. The treatment of stars with Z $<$ 0.002 is however only important for our predictions at Galactic ages smaller than a few Gyr, as afterwards the number of such low Z systems merging is negligible. To illustrate, in our model 2 they make up less than 5\% of the total merger rate after 3.5 Gyr. Systems merging at present (13.2 Gyr) have average metallicities of 0.014 (BH+NS) to 0.016 (NS+NS).}. It then continuously rises with time, as enrichment is provided by stellar ejecta. The Fe-evolution of a galaxy during the first 100 Myr is determined by the massive stars but afterwards one of the most important contributions is the Fe-enrichment by type Ia supernovae (SNe Ia). Therefore, it is important to include these events as best as possible. The most obvious way is to include the SN Ia rate from observations, the so called delay time distribution (DTD). However, since this observed DTD is not available for different values of Z, it does not allow to judge how strongly it depends on metallicity. Using the same DTD at all Z may thus only prove to be wrong. Therefore, we prefer to include the SN Ia rate we have self-consistenly calculated in detail, including its Z-dependence. Specifically, we use the best model from Mennekens et al. (2010) which for solar Z indeed reproduces the observed DTD (but which is very different for low Z). Once this method of including SN Ia rates has been selected, the resulting evolution of [Fe/H] is always very similar. This is true regardless of the exact value of many other galactic evolutionary parameters about which there exist significant uncertainties (e.g. the effect of radial migration). But since the metallicity evolution is the only galactic property significantly influencing DCO merger rates, these uncertainties do not play an important role in this study. The metallicity evolution simulations are illustrated in Fig. \ref{fig:AMR} for the two SFR models used in the present paper and they can be considered as typical for all the calculations of the present paper.

i.	Korobkin et al. (2012) calculated the r-process ejecta of NS+NS and BH+NS (or NS+BH) binary mergers. They are implemented in our galactic code in order to simulate the temporal evolution of the r-process elements in the Milky Way.

\section{Results}

First remind that all our simulations refer to a galactic star formation rate that is either constant in time (SFR$_2$) or satisfies the function illustrated in Fig. \ref{fig:SFR} (SFR$_1$). Moreover, although the age-metallicity relation is calculated self-consistently (remember that we only consider the models for the temporal evolution of the SNIa rate that also fit the observed DTD of elliptical galaxies as explained in the previous section), it is interesting to notice that in all our simulations this relation very closely matches one of those given in Fig. \ref{fig:AMR}. Table \ref{tab:models} lists the parameters of various models for which the results are given in the present paper. All of these have been discussed in Sect. 2 and 3. The SFR index refers to those defined there and shown in Fig. \ref{fig:SFR}. The presence or absence of ECSNe, the LBV scenario and BH kicks are as discussed. So are the used fall back models, average NS and (when not turned off) BH kick velocities and initial-final mass relations. The values of $\alpha_{CE}$ and $\beta$ refer to the process of RLOF/common envelope/spiral-in of massive binaries. The `5' indicates the model where the value of $\lambda$ is multiplied by that factor. The period distribution labeled `alternative' means the period distribution proposed by Sana et al. (2013). Angular momentum loss is either that of the second Lagrangian point (L$_2$) or gainer orbital angular momentum loss (`O').

\begin{table*}
\centering
\caption{The different models for which simulations are computed for the present paper. See text for the definition of parameters and their values.}
\begin{tabular}{c c c c c c c c c c c c}
\hline
Model &	SFR &	ECSN & $\beta$ & $\alpha_{CE}$ & Fall back & Period & LBV & BH & avg. NS & $M_i-M_f$ & AM \\
 & & & & & model & distribution & & kicks & kick (km/s) & rel. & loss \\
\hline
1 & 2 & Off & 1 & 0.5 & Rapid & Flat & On & On & 450 & Brussels & $L_2$ \\
2 & 2 & On & 1 & 0.5 & Rapid & Flat & On & On & 450 & Brussels & $L_2$ \\
3 & 1 & Off & 1 & 0.5 & Rapid & Flat & On & On & 450 & Brussels & $L_2$ \\
4 & 1 & On & 1 & 0.5 & Rapid & Flat & On & On & 450 & Brussels & $L_2$ \\
5 & 1 & On & 1 & 0.5 & Delayed & Flat & On & On & 450 & Brussels & $L_2$ \\
6 & 2 & On & 1 & 1 & Rapid & Flat & On & On & 450 & Brussels & $L_2$ \\
7 & 2 & Off & 1 & 1 & Rapid & Flat & On & On & 450 & Brussels & $L_2$ \\
8 & 2 & On & 1 & 0.1 & Rapid & Flat & On & On & 450 & Brussels & $L_2$ \\
9 & 2 & On & 0.5 & 0.5 & Rapid & Flat & On & On & 450 & Brussels & $L_2$ \\
10 & 2 & On & 0.1 & 0.5 & Rapid & Flat & On & On & 450 & Brussels & $L_2$ \\
11 & 1 & On & 0.1 & 1 & Rapid & Flat & On & On & 450 & Brussels & $L_2$ \\
12 & 2 & On & 1 & 0.5 & Rapid & Flat & Off & On & 450 & Brussels & $L_2$ \\
13 & 2 & On & 1 & 0.5 & Rapid & Alternative & On & On & 450 & Brussels & $L_2$ \\
14 & 2 & On & 1 & 0.5 & Rapid & Flat & On & Off & 450 & Brussels & $L_2$ \\
15 & 2 & On & 1 & 0.5 & Rapid & Flat & On & Off & 265 & Brussels & $L_2$ \\
16 & 2 & On & 1 & 1 & Rapid & Flat & On & On & 450 & Fryer & $L_2$ \\
17 & 2 & On & 0.5 & 0.5 & Rapid & Flat & Off & On & 450 & Brussels & $L_2$ \\
18 & 2 & On & 0.1 & 0.5 & Rapid & Flat & Off & On & 450 & Brussels & $L_2$ \\
19 & 2 & On & 1 & `5' & Rapid & Flat & On & On & 450 & Brussels & $L_2$ \\
20 & 2 & On & 1 & 1 & Rapid & Flat & Off & On & 450 & Brussels & $L_2$ \\
21 & 2 & On & 0.5 & 1 & Rapid & Flat & On & On & 450 & Brussels & `O' \\
22 & 2 & On & 0.5 & 1 & Rapid & Flat & On & On & 450 & Brussels & $L_2$ \\
23 & 2 & On & 0.5 & 1 & Rapid & Flat & Off & On & 450 & Fryer & `O' \\
\hline
\end{tabular}
\label{tab:models}
\end{table*}

\begin{figure*}[t]
\centering
   \includegraphics[width=14cm]{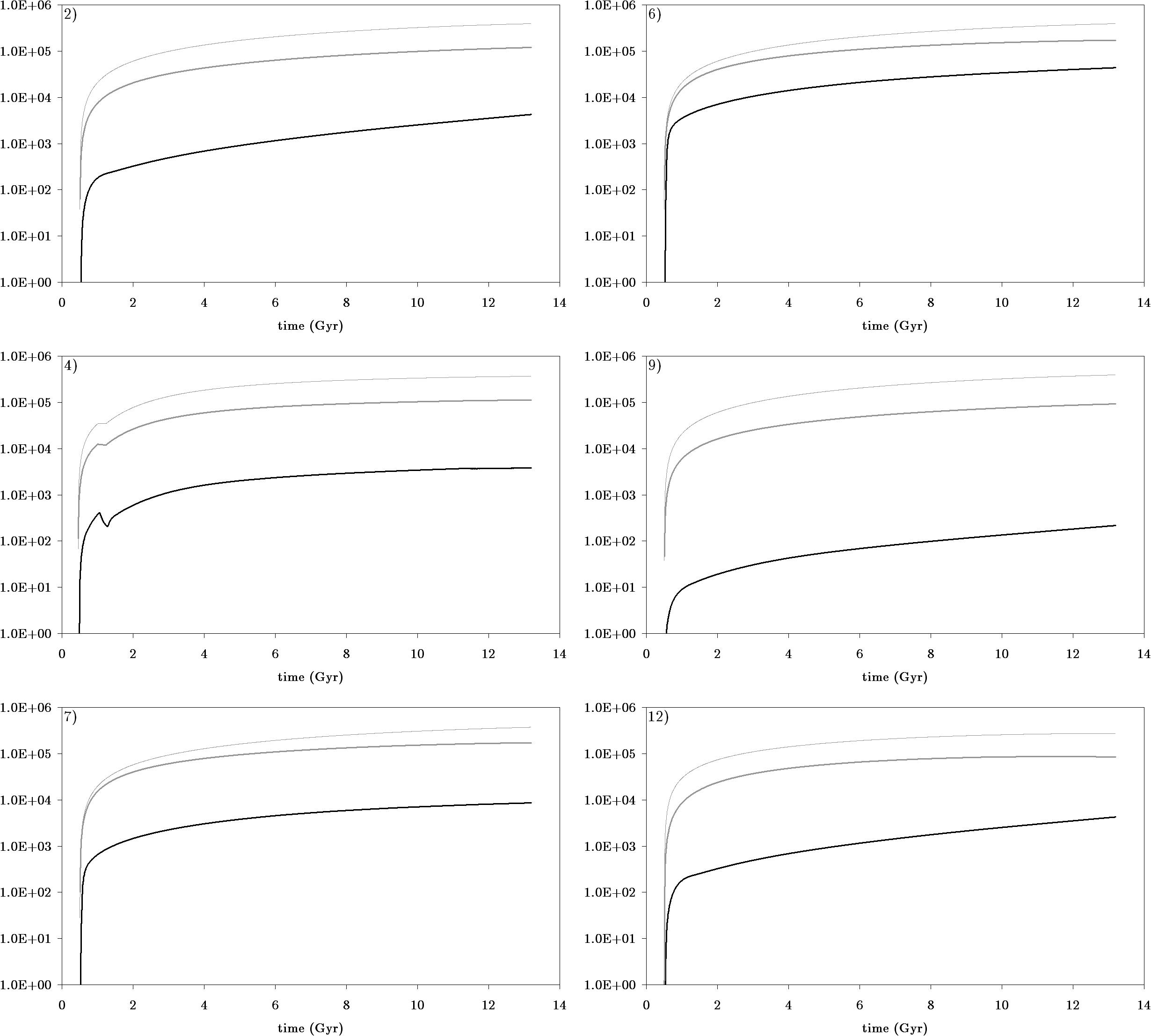}
     \caption{Number of galactic NS+NS (solid black), combined BH+NS and NS+BH (solid gray) and BH+BH (thin gray) binaries. Corresponding model numbers are indicated in the top left corner of all figure panels.}
     \label{fig:Pop}
\end{figure*}

\begin{figure*}[t]
\centering
   \includegraphics[width=14cm]{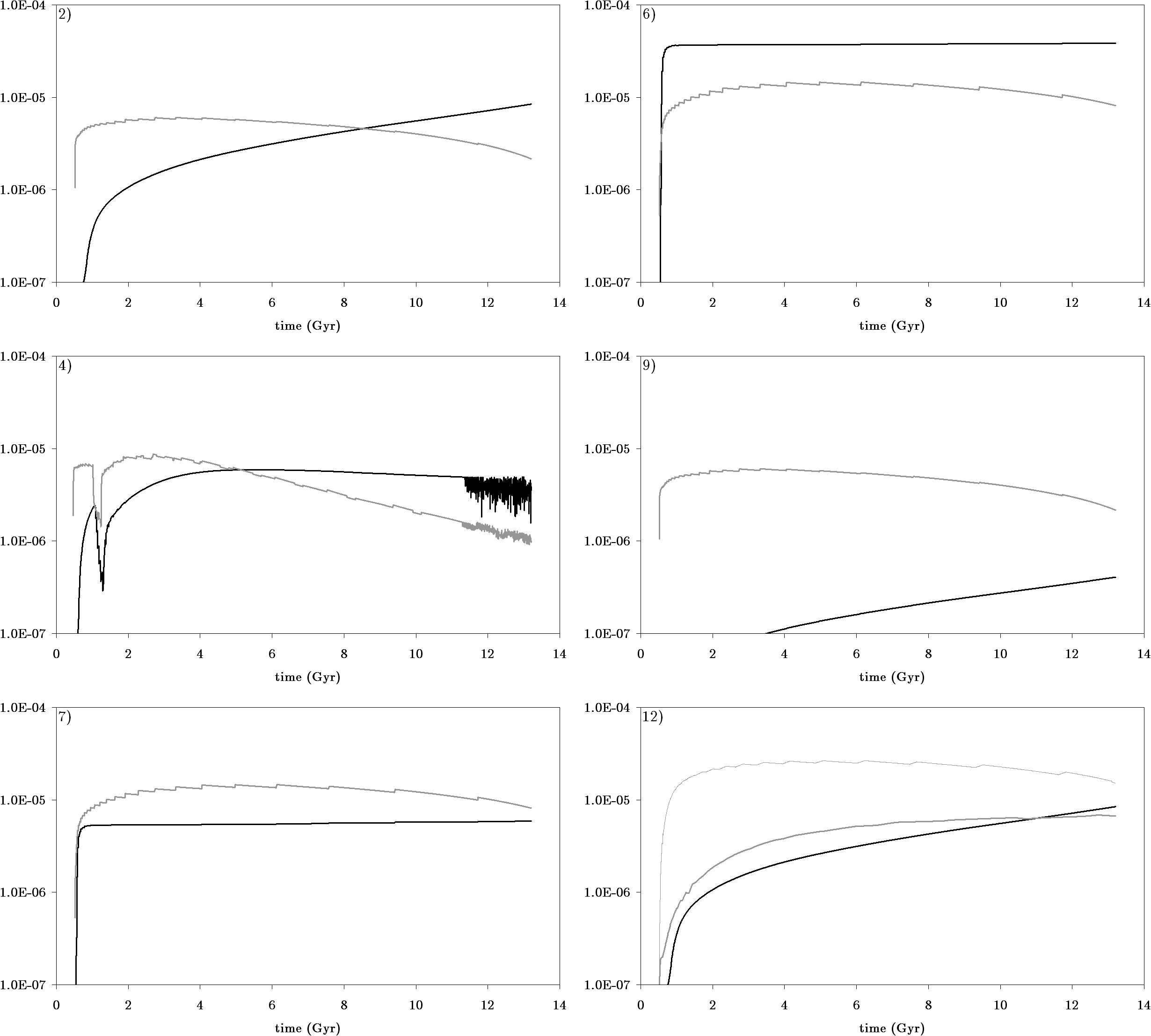}
     \caption{Galactic double compact binary merger rates per year, seperated into double NS mergers (black), mixed BH+NS mergers (solid gray) and (when applicable) double BH mergers (thin gray). Corresponding model numbers are indicated in the top left corner of all figure panels.}
     \label{fig:mergers}
\end{figure*}

\begin{figure*}[t]
\centering
   \includegraphics[width=14cm]{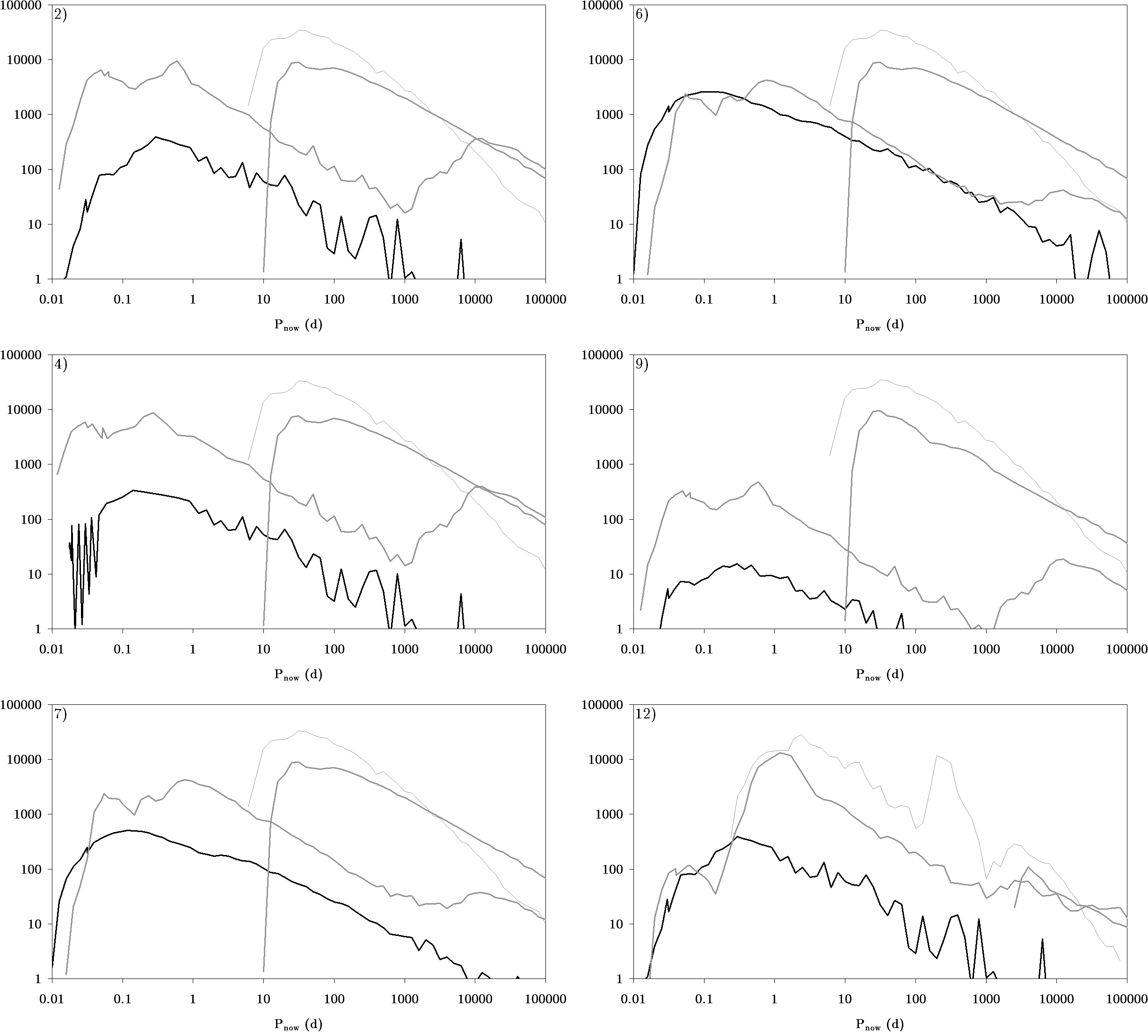}
     \caption{Theoretically expected period distributions at present of the galactic NS+NS (solid black), BH+NS (leftmost solid gray), NS+BH (rightmost solid gray) and BH+BH (thin gray) binaries. Corresponding model numbers are indicated in the top left corner of all figure panels.}
     \label{fig:Pdistr}
\end{figure*}

For a few typical models defined in Table \ref{tab:models} we show in Fig. \ref{fig:Pop} the temporal evolution of the number of galactic NS+NS, BH+NS, NS+BH and BH+BH binaries. Fig. \ref{fig:mergers} then illustrates the temporal evolution of the galactic double compact binary merger rates. The rates at present are then translated into LIGO II rates using the method and parameter values outlined by Belczynski \& Dominik (2012), which is very similar to e.g. the procedure by Sadowski et al. (2008). This detection horizon is taken from the official LIGO website https://www.advancedligo.mit.edu/science.html. A slightly lower value would obviously not change the conclusions significantly, but merely require the downscaling of the predicted rates with (the third power of) the factor with which the horizon is reduced. The results for the models of Table \ref{tab:models} are given in Table \ref{tab:results}. The theoretically expected period distributions at present of the galactic NS+NS, BH+NS, NS+BH and BH+BH binaries are given in Fig. \ref{fig:Pdistr}. All these figures are presented for six different models from Table \ref{tab:models}: three which represent our most standard assumptions and match the observational constraints on NS+NS merger rates well\footnote{Interestingly, we can do a similar exercise as was done by Dominik et al., e.g. we can test our simulations by comparing the predicted Galactic double NS merger rate with the observed lower limit (3 Myr$^{-1}$) proposed by Kim et al. (2010). This value is not met by several models from Table \ref{tab:results} and they may therefore be ruled out.}, two which predict a very high, respectively low number of NS+NS mergers, and one with the LBV scenario off (and thus a contribution from BH+BH mergers).

\begin{figure*}[t]
\centering
   \includegraphics[width=14cm]{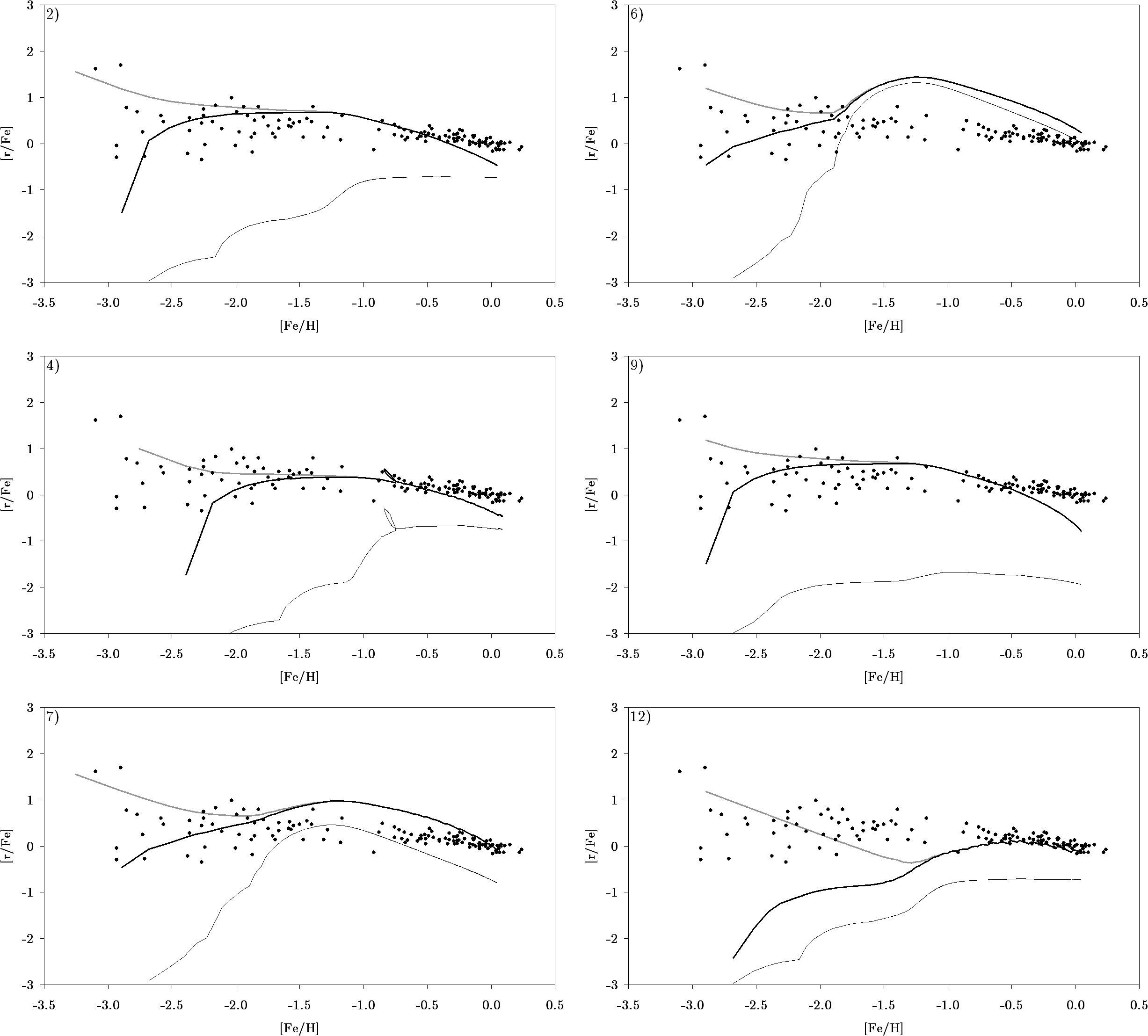}
     \caption{Predicted (solid black) and observed (dots) temporal evolution of the r-process element yields, as well as predicted evolution with double NS mergers only (thin black) and with core collapse supernova yields included (solid gray, see text). Corresponding model numbers are indicated in the top left corner of all figure panels.}
     \label{fig:r}
\end{figure*}

\begin{table*}
\centering
\caption{The predicted LIGO II rates, the predicted absolute galactic merger rate, the predicted number of double compact star binaries in the Solar neighborhood. The result of the mixed systems is always the sum of the NS+BH and BH+NS binaries.}
\begin{tabular}{c c c c c c c c c c}
\hline
 & \multicolumn{3}{c}{LIGO II rates (yr$^{-1}$)} & \multicolumn{3}{c}{Galactic merger rates (Myr$^{-1}$)} & \multicolumn{3}{c}{Solar neighborhood population} \\
Model & NSNS & BHNS & BHBH & NSNS & BHNS & BHBH & NSNS & BHNS & BHBH \\
\hline
1 & 0.39 & 28.6 & 0 & 1.27 & 2.16 & 0 & 0.66 & 120 & 372 \\
2 & 2.55 & 28.6 & 0 & 8.43 & 2.16 & 0 & 4.26 & 120 & 393 \\
3 & 0.16 & 12.7 & 0 & 0.54 & 0.98 & 0 & 0.59 & 113 & 348 \\
4 & 1.07 & 12.7 & 0 & 3.54 & 0.98 & 0 & 3.82 & 113 & 367 \\
5 & 1.07 & 18.5 & 0 & 3.54 & 1.31 & 0 & 3.71 & 119 & 355 \\
6 & 12.6 & 101 & 0 & 38.4 & 8.15 & 0 & 43.8 & 172 & 393 \\
7 & 1.96 & 101 & 0 & 5.87 & 8.15 & 0 & 8.62 & 172 & 372 \\
8 & 0 & 0.04 & 0 & 0 & 0.005 & 0 & 0 & 116 & 393 \\
9 & 0.13 & 28.6 & 0 & 0.41 & 2.16 & 0 & 0.22 & 92.7 & 393 \\
10 & 0.04 & 28.6 & 0 & 0.11 & 2.16 & 0 & 0.06 & 89.2 & 393 \\
11 & 0.14 & 57.7 & 0 & 0.39 & 4.6 & 0 & 1.0 & 125 & 367 \\
12 & 2.55 & 87 & 4780 & 8.43 & 6.33 & 15.6 & 4.26 & 84 & 273 \\
13 & 1.86 & 23 & 0 & 6.14 & 1.73 & 0 & 3.12 & 235 & 468 \\
14 & 2.55 & 157 & 0 & 8.43 & 16.3 & 0 & 4.26 & 151 & 2020 \\
15 & 4.23 & 187 & 0 & 13.9 & 19.3 & 0 & 8.80 & 258 & 2020 \\
16 & 10.2 & 38.8 & 7140 & 33.8 & 5.30 & 97.3 & 39.9 & 58.0 & 657 \\
17 & 0.13 & 99.2 & 3810 & 0.41 & 7.30 & 11.6 & 0.22 & 84 & 164 \\
18 & 0.04 & 94.9 & 2980 & 0.11 & 6.97 & 8.23 & 0.06 & 89 & 128 \\
19 & 49.6 & 150 & 0 & 153 & 12.7 & 0 & 1290 & 391 & 393 \\
20 & 12.6 & 67.3 & 4190 & 38.4 & 5.42 & 13.8 & 43.8 & 159 & 392 \\
21 & 11.7 & 106 & 0 & 36.2 & 8.49 & 0 & 24.3 & 150 & 395 \\
22 & 0.85 & 101 & 0 & 2.49 & 8.09 & 0 & 3.52 & 144 & 393 \\
23 & 1.94 & 484 & 10100 & 7.43 & 68.2 & 76.6 & 2.36 & 334 & 2680 \\
\hline
\end{tabular}
\label{tab:results}
\end{table*}

Dominik et al. (2012) thoroughly discussed the effects of different binary evolutionary parameters on the predicted population of double compact star binaries and the merger rates. Here we mainly focus on differences with our results. Before discussing the differences it may be worth mentioning also an encouraging similarity. Most of the double NS mergers result from the initial mass range below 20 M$_{\odot}$ so that differences between StarTrack and the Brussels code for larger initial masses should not be very important if double NS merger rates are compared. As far as the physics and the chosen parameter values are concerned the model V12 of Dominik et al. (2012) (corresponding to conservative mass transfer during stable RLOF, $\beta$ = 1) and our model 6 should be similar and this is indeed the case. V12 predicts a present day Galactic NS-NS merger rate of 47 per Myr whereas our model 6 yields 38 per Myr. This similarity is remarkable accounting for the fact that StarTrack uses a very crude Galactic model whereas the Brussels code does not account for tidal effect in binaries.

The effect of $\beta$ on our results can be illustrated by comparing the models 2, 9 and 10. The smaller $\beta$ the smaller the number of double NS binaries. Compared to $\beta$ = 1, $\beta$ = 0.5 predicts a double NS merger rate that is a factor 20 smaller. This is quite different when compared to the predictions made by Dominik et al. We think that the difference is mainly due to the difference in the model of angular momentum loss during a non-conservative RLOF. As discussed in Sect. 2 we use a model where mass leaves the binary via the second Lagrangian point L$_2$ and forms circumbinary disk. Mass that leaves the binary therefore implies a large orbital angular momentum loss and when $\beta$ = 0.5 many case Br binaries merge. Dominik et al. use a model where mass leaves the binary essentially taking with the specific angular momentum of the gainer. This is a much smaller angular momentum loss than with our L$_2$ model. To illustrate, we implemented the model of StarTrack in our code and the results are given in Table \ref{tab:results} as model 21. As can be noticed we now have a significantly larger double NS merger rate more in line (within a factor 2) with the standard model in Dominik et al.  This experiment illustrates the very large effect of the model to describe the angular momentum loss during a non-conservative (stable) RLOF. Since there is no consensus at present the difference between model 21 and 22 (a factor 20-40) may be representative for the uncertainty of population synthesis predictions.

In all the models with LBV/RSG scenario on and with our preferred initial-final mass relation most of the double BH binaries have very large periods and they never merge (confirming the results of De Donder et al., 1997 and De Donder and Vanbeveren, 2004). We therefore predict no LIGO II detections from these systems, a conclusion which is significantly different from the one of Voss and Tauris (2003) and of Dominik et al. (2012) who predict that the LIGO II detection rate will be dominated by merging double BH binaries. This very large difference can be attributed entirely to the adopted LBV scenario for binaries with component masses $\geq$ 40 M$_{\odot}$, the RSG scenario for binaries with component masses smaller than 40 M$_{\odot}$ and our initial mass-final mass relation (see Sect. 2). To illustrate we switched the LBV and the RSG scenario off and we also adopted a classical RLOF/common envelope/spiral-in phase for these binaries. The results are given in Table \ref{tab:results}, model 12. As can be noticed the expected double BH merger (and LIGO II) rate is now by far the largest, and more in line with the results of Voss and Tauris and of Dominik et al. In model 12 we still adopt our initial-final mass relation. Model 16 then is similar as model 12 but with the single star initial-final mass relation of Fryer et al. (2012) used by Dominik et al. (however without their adaptations to include binary effects as these are not entirely clear to us, see also Sect. 2). As expected the double BH merger and LIGO II rates become even more dominant. Model 23 shows a calculation of ours which approximates the standard model of Dominik et al. (2012) as much as possible. It is interesting to note that if we incorporate the Fryer et al. initial-final mass relation for single stars, we find results that are similar\footnote{The only significant difference is in the number of mixed systems, which is larger in our simulation. This is due to the fact that these originate from systems with a small initial mass ratio, which are much more likely to merge with the StarTrack criterion for stable mass transfer than with ours (see Sect. 2.2).} to those of Dominik et al. (2012) and Belczynski \& Dominik (2012). \emph{Therefore, when LIGO II will be operational}, the observed rates may help to decide upon the effects of LBV/RSG mass loss and/or of the initial-final mass relation of massive close binaries.

Our population number synthesis simulations let us suspect that NS+BH LIGO II rates will outnumber the double NS rates, confirming the results of Dominik et al. (2012). Furthermore, the overall predictions hardly depend on the adopted period distribution or on the adopted fall-back scenario (see models 5 and 13). Comparison of models 2, 14 and 15 allows to test the importance of the assumed NS kick velocities and the presence or absence of BH kicks. While these assumptions (sometimes significantly) affect the exact rates, they are not critical to our conclusions.

We now study the influence on our results of metallicity and SFR\footnote{During the reviewing process of the present work, a paper by Dominik et al. (2013) appeared which treats the cosmological implications for the detection rates in more detail. By using a star formation history, galaxy mass distribution and galaxy redshift-metallicity relation, they predict merger rates as a function of redshift. Studies such as ours implicitly assume that the current evolutionary stage of the Milky Way is typical for all galaxies in the detection region, whereas Belczynski \& Dominik (2012) use the simple procedure outlined in Sect. 3 to approximate the local stellar content of the universe. Dominik et al. (2013) in contrast find that the presence of local low-metallicity, high star forming galaxies may mean that at low redshift, the relative merger rates for the different types of DCOs may be very different from their Galactic results. While taking these effects into account would obviously result in a more detailed determination of detection rates, this would not change our final conclusions about the importance of the initial-final mass relation and LBV and RSG wind mass loss. Nor does it have implications for our Galactic chemical predictions.}. Recall that these evolve through time as shown in Fig. \ref{fig:AMR} and \ref{fig:SFR}. Fig. \ref{fig:mergers} presents the Galactic merger rates from t=0 on, and hence also includes stars formed at very low metallicity and (in the case of SFR$_1$) very high star formation. Nevertheless, the BH+BH merger rate is still zero in all the models with our own initial-final mass relation and the LBV scenario on. The reason is that (as is visible from Fig. \ref{fig:MiMf}), our initial-final mass relation is relatively Z-independent, and the same is true for the LBV mass loss rates. To further illustrate, we can compare the merger rates obtained for starburst galaxies with metallicities of respectively 0.02 and 0.002. For the low metallicity, we find that BH+NS LIGO detection rates increase by a factor of three (in line with the findings of Belczynski et al. 2010a), while NS+NS detection rates drop by a factor of eight, compared to Solar metallicity. Double BH merger rates remain zero. For a galaxy with high SFR, it is obvious that all merger rates come to lie higher. However, in our case, the zero BH+BH merger rate remains zero, nor is the relative amount of NS+NS vs. BH+NS mergers affected.

Since the binary population code is linked to a galactic chemical evolutionary code, by using the r-process yields of Korobkin et al. (2012) it is straightforward to calculate the temporal evolution of the r-process elements produced during the merging of double NS and merging BH+NS or NS+BH binaries. Fig. \ref{fig:r} shows the predicted and the observed temporal evolution of the yields (we focus on europium)\footnote{The observations are taken from Woolf et al. (1995), Burris et al. (2000) and Chen et al. (2006).}. We separated the contribution of double NS binaries and the one of BH+NS systems (with the LBV scenario on, the NS+BH binaries are expected to have rather large periods and merging within a Hubble time is rather unlikely; they therefore do not significantly contribute to the r-process enrichment).

The results indicate that except for the earliest phase ([Fe/H] $< -2.5$) double compact star mergers contribute significantly to the chemical enrichment of r-process elements of the Galaxy. This enrichment is mainly due to the BH+NS binary mergers, the double NS binary mergers (and the NS+BH mergers) do only marginally contribute. Furthermore, while recent results (e.g. Banerjee et al., 2011; Qian, 2012) suggest that, exactly and only at these very low metallicities, r-process element enrichment may also be provided by core collapse SNe, the discrepancy for [Fe/H] $< -2.5$ can also be relaxed by using a chemical evolution model more in the line of the one suggested by Cescutti et al. (2013) for the halo.

We have also calculated our models including this r-process element production by type II core collapse supernovae, however as argued by Banerjee et al. (2011) only at Z $< 0.001$. For the europium yield of a single event we take $10^{-8}$ M$_{\odot}$, an average of the values published by e.g. Argast et al. (2004) and Cescutti et al. (2006). Also following Argast et al. (2004), we assume that 1.2\% of r-process elements consist of europium. The thus obtained results are also shown in Fig. \ref{fig:r}. It is obvious that this order-of-magnitude estimate does indeed cause the [r/Fe]-values to lie significantly higher for [Fe/H] $< -2$, but does not influence them anymore afterward.\footnote{This paragraph was added during the reviewing process of the present work, after similar conclusions were reached by Matteucci et al. (2014).}

\section{Summary}

In the present paper we discussed the merger rates of double NS binaries, double BH binaries and of mixed NS+BH systems expected by binary population synthesis models and we translated the predictions into LIGO II rates. We first compared the initial-final mass relations used in different population codes and the effects on predicted rates. We then highlighted the importance of Luminous Blue Variable stellar wind mass loss on the evolution of close binaries with component masses $\geq$ 40 M$_{\odot}$ and of the red supergiant mass loss on the evolution of close binaries with component masses $\leq$ 40 M$_{\odot}$. A most important conclusion is that with our preferred model of massive close binary evolution we do not expect many double BH mergers contrary to simulations done in the recent past by other research teams. Furthermore, since the predicted population of double NS binaries depends critically on assumptions related to the process of stable RLOF and/or the efficiency for converting orbital energy into escape energy during the common envelope of a binary, by comparing with the observed population we have to rule out models where the stable RLOF is highly non-conservative and/or the common envelope efficiency is very low.

By linking a binary population code, a galactic code and the r-process ejecta of double compact star mergers that have recently become available, it is possible to compute in detail the galactic temporal evolution of these ejecta and compare with observation. We conclude that (with exception for the first say 100 Myr) compact star binary mergers are major contributors to the r-process enrichment. Most of this enrichment is due to BH+NS mergers, the contribution of the NS+NS mergers is marginal.

\begin{acknowledgements}
      We thank Gabriele Cescutti and an anonymous referee for constructive comments that helped to improve the manuscript.
\end{acknowledgements}

\end{document}